\documentclass[10pt,conference]{IEEEtran}
\usepackage{amsmath,amssymb,amsthm} 
\usepackage[dvips]{graphicx}
\usepackage{cite}
\usepackage{balance}
\newtheorem{example}{Example}
\newtheorem{proposition}{Proposition}
\newcommand{\argmin}{\mathop{\mathrm{argmin}}\limits} 
\newcounter{MYtempeqncnt}

\begin{document}

% paper title
%\title{Instructions for Preparation and Submission of Papers for the 
%Proceedings of ISIT 2009}
\title{Large-System Analysis of Joint User Selection\\ and Vector Precoding  
with Zero-Forcing Transmit Beamforming for MIMO Broadcast Channels}

\author{
  \IEEEauthorblockN{Keigo Takeuchi}
  \IEEEauthorblockA{
    Dept.\ Commun.\ Engineering \& Inf.\\ 
    University of Electro-Communications\\
    Tokyo 182-8585, Japan\\
    Email: ktakeuchi@uec.ac.jp
  }
  \and
  \IEEEauthorblockN{Ralf R. M\"uller}
  \IEEEauthorblockA{
    Dept Elec.\ \& Telecommun.\\ 
    NTNU\\
    NO--7491 Trondheim, Norway\\
    Email: ralf@iet.ntnu.no
  }
  \and
  \IEEEauthorblockN{Tsutomu Kawabata}
  \IEEEauthorblockA{
    Dept.\ Commun.\ Engineering \& Inf.\\ 
    University of Electro-Communications\\
    Tokyo 182-8585, Japan\\
    Email: kawabata@uec.ac.jp
  }
%  \IEEEauthorblockN{Ralf R. M\"uller}
%  \IEEEauthorblockA{
%    Dept Elec. \& Telecommun.\\ 
%    NTNU\\
%    NO--7491 Trondheim, Norway\\
%    Email: ralf@iet.ntnu.no
%  }
%  \and
%  \IEEEauthorblockN{Mikko Vehkaper\"a}
%  \IEEEauthorblockA{
%    Dept Elec. \& Telecommun.\\ 
%    NTNU\\
%    NO--7491 Trondheim, Norway\\
%    Email: mikko@iet.ntnu.no
%  }
}

%\author{\IEEEauthorblockN{Keigo Takeuchi\IEEEauthorrefmark{1},
%Ralf R. M\"uller\IEEEauthorrefmark{2}, 
%Mikko Vehkaper\"a\IEEEauthorrefmark{2}, and 
%Toshiyuki Tanaka\IEEEauthorrefmark{3}}
%\IEEEauthorblockA{\IEEEauthorrefmark{1}Dept Information \& 
%Communication Engineering, the University of Electro-Communications, 
%Tokyo 182-8585, Japan}
%\IEEEauthorblockA{\IEEEauthorrefmark{2}Dept Elec. \& 
%Telecommun., Norwegian University of Science and Technology, 
%NO--7491 Trondheim, Norway}
%\IEEEauthorblockA{\IEEEauthorrefmark{3}Graduate School of Informatics, 
%Kyoto University, Kyoto 606--8501, Japan \\
%Email: takeuchi@ice.uec.ac.jp, mikko@iet.ntnu.no, ralf@iet.ntnu.no, 
%tt@i.kyoto-u.ac.jp} 
%}

% make the title area
\maketitle

\begin{abstract}
Multiple-input multiple-output (MIMO) broadcast channels (BCs) 
(MIMO-BCs) with perfect channel state information (CSI) at the transmitter are 
considered. As joint user selection (US) and vector precoding (VP) (US-VP) 
with zero-forcing transmit beamforming (ZF-BF), US and continuous VP (CVP) 
(US-CVP) and data-dependent US (DD-US) are investigated. The replica method, 
developed in statistical physics, is used to analyze the energy penalties for 
the two US-VP schemes in the large-system limit, where the number of users, 
the number of selected users, and the number of transmit antennas tend to 
infinity with their ratios kept constant. Four observations are obtained in 
the large-system limit: First, the assumptions of replica symmetry (RS) and 
1-step replica symmetry breaking (1RSB) for DD-US can provide acceptable 
approximations for low and moderate system loads, 
respectively. Secondly, DD-US outperforms CVP with random US in terms of the 
energy penalty for low-to-moderate system loads. Thirdly, the asymptotic 
energy penalty of DD-US is indistinguishable from that of US-CVP for low 
system loads. Finally, a greedy algorithm of DD-US proposed in authors' 
previous work can achieve nearly optimal performance for low-to-moderate 
system loads.  
\end{abstract}

\section{Introduction}
Multiple-input multiple-output (MIMO) broadcast channels (BCs) (MIMO-BCs) are 
a model for the downlink of multiuser MIMO systems. The capacity region of the 
MIMO-BCs with perfect channel state information (CSI) at the transmitter has 
been shown to be achieved by dirty-paper coding 
(DPC)~\cite{Caire03,Viswanath03,Yu04,Weingarten06}, which is a coding scheme 
to {\em pre-cancel} inter-user interference at the transmitter 
side~\cite{Costa83}. Since DPC is infeasible in terms of complexity, it is 
an important research topic to construct a suboptimal scheme that can achieve 
an acceptable tradeoff between performance and complexity.    

Zero-forcing transmit beamforming (ZF-BF) is a naive approach for eliminating 
inter-user interference at the transmitter~\cite{Spencer04,Wiesel08}. 
A drawback of ZF-BF is that the energy required for pre-cancellation of 
inter-user interference, called 
{\em energy penalty} in this paper, diverges as the system load increases. 
An increase of the energy penalty results in a degradation of the receive 
signal-to-noise ratio (SNR). 

User selection (US) with ZF-BF~\cite{Dimic05,Yoo06} is a 
promising approach when the number of users is much greater than the number of 
transmit antennas. Interestingly, it was proved 
that a greedy algorithm for US with ZF-BF can achieve the sum capacity of the 
MIMO-BC when {\em only} the number of users tends to 
infinity~\cite{Yoo06,Wang08}. This result is because it is possible to select 
a finite subset of users who have almost orthogonal channel vectors in that 
limit. As the number of transmit antennas increases, it becomes difficult to 
select such a subset of users, since the number of selected users should be 
increased to improve the throughput. This implies that US with ZF-BF is 
suboptimal in the situation where the number of transmit antennas is 
comparable to the number of users. Such a situation is becoming 
practical~\cite{Marzetta10}. The goal of this paper is to construct a 
precoding scheme that works well in that situation. 

Vector perturbation~\cite{Hochwald05} or vector precoding 
(VP)~\cite{Mueller08} is a sophisticated precoding scheme suited for the 
situation where the number of transmit antennas is comparable to the number of 
users. In VP, the data vector is modified 
to take values in a relaxed alphabet~\cite{Hochwald05,Mueller08}. 
This relaxation reduces the energy penalty without degrading the minimum 
distance between the data symbols. As relaxed alphabets, a lattice-type 
alphabet~\cite{Hochwald05,Razi10} and a continuous alphabet~\cite{Mueller08} 
were proposed. In this paper, VP schemes with the lattice-type and continuous 
alphabets are referred to as ``lattice VP (LVP)'' and ``continuous VP 
(CVP),'' respectively.  The search for a vector to minimize the energy penalty 
reduces to a convex optimization problem\footnote{
The problem is non-convex for joint user selection and CVP considered in 
this paper.} for CVP, while the search problem for LVP is NP-hard. 
However, CVP might be still hard to implement since the convex optimization 
has to be solved every time slot. The goal of our research is to propose a 
more practical precoding scheme. 

We propose joint US and VP (US-VP), and analyze the performance in the 
large-system limit, in which the number of transmit antennas $N$, the number 
of users $K$, and the number of selected users $\tilde{K}$ tend to infinity 
with the ratios $\alpha=K/N$ and $\kappa=\tilde{K}/K$ kept constant. 
In this paper, $\alpha\kappa=\tilde{K}/N$ is referred to as the system load. 

Data-dependent US (DD-US) proposed in \cite{Takeuchi121} is regarded as a 
special case of US-VP, i.e.\ as US-VP with the original alphabet as the 
relaxed alphabet. DD-US takes into account the data symbols, along with the 
channel vectors, to reduce the energy penalty, as VP does. Furthermore, 
DD-US can be implemented with a suboptimal greedy algorithm~\cite{Takeuchi121}. 

%Combining US and VP is a natural idea and the main topic in this paper. 
%Razi et al.~\cite{Razi10} investigated a separated scheme of US and VP, in 
%which US is performed before VP. In the separated scheme, the selected subset 
%of users is independent of the data symbols. Processing US and VP 
%simultaneously may achieve smaller energy penalty than the separated method. 

%The main contribution of this paper is to show that a practical greedy 
%algorithm for DD-US proposed in \cite{Takeuchi121} can achieve the 
%{\em minimum} energy penalty for low-to-moderate system loads in the 
%large-system limit. Furthermore, the minimum energy penalty for DD-US is 
%shown to be indistinguishable from those for US-LVP and US-CVP in the same 
%load regime. 

The large-system analysis presented in this paper is based on the 
{\em non-rigorous} replica method, developed in statistical 
physics~\cite{Mezard87,Nishimori01}. The replica method is a powerful tool for 
analyzing the large-system performance of MIMO 
systems~\cite{Tanaka02,Mueller08,Zaidel12}. 
Several results based on the replica method have been justified rigorously. 
See \cite{Korada10,Guerra032,Talagrand06} for the details.

\section{MIMO Broadcast Channel} 
We consider a Gaussian MIMO-BC with perfect CSI at the 
transmitter, which consists of a base station with $N$ transmit 
antennas and $K$ receivers (users) with one receive antenna. The coherence 
time $T_{\mathrm{c}}$ is assumed to be sufficiently long\footnote{
The base station may utilize the reciprocal channel to obtain CSI in practice. 
In order to attain accurate CSI, the coherence time $T_{\mathrm{c}}$ should be 
at least larger than the number of users $K$. In this paper, the limit 
$T_{\mathrm{c}}\rightarrow\infty$ is implicitly taken before the large-system 
limit.}. 
Let $y_{k,t}\in\mathbb{C}$ denote the received signal for user~$k$ in time 
slot~$t$ ($t=0,1,\ldots,T_{\mathrm{c}}-1$). The received vector 
$\boldsymbol{y}_{t}=(y_{1,t},\ldots,y_{K,t})^{\mathrm{T}}$ in time slot~$t$ 
is given by 
\begin{equation} \label{MIMO-BC} 
\boldsymbol{y}_{t} = \frac{1}{\sqrt{\mathcal{E}}}
\boldsymbol{H}\boldsymbol{u}_{t} + \boldsymbol{n}_{t}, 
\quad \boldsymbol{n}_{t}
\sim\mathcal{CN}(\boldsymbol{0},N_{0}\boldsymbol{I}_{K}).  
\end{equation}
In (\ref{MIMO-BC}), 
$\boldsymbol{u}_{t}=(u_{1,t},\ldots,u_{N,t})^{\mathrm{T}}\in\mathbb{C}^{N}$ 
denotes a transmit vector in time slot~$t$, defined in the next section. 
The $(k,n)$-element of the channel matrix 
$\boldsymbol{H}\in\mathbb{C}^{K\times N}$ represents a complex channel gain 
between the $n$th transmit antenna and the $k$th user. 
The energy penalty $\mathcal{E}$ is defined as 
the time average of the power of the transmit vectors 
\begin{equation} \label{energy_penalty} 
\mathcal{E}=\frac{1}{T_{\mathrm{c}}}
\sum_{t=0}^{T_{\mathrm{c}}-1}\|\boldsymbol{u}_{t}\|^{2}. 
\end{equation}
Thus, the prefactor $\mathcal{E}^{-1/2}$ 
in (\ref{MIMO-BC}) implies that the transmit SNR is constrained to $1/N_{0}$.  

We assume that $\boldsymbol{H}$ is known to the 
transmitter, and that $\boldsymbol{H}$ has mutually independent circularly 
symmetric complex Gaussian entries with variance $1/N$. These idealized 
assumptions allow us to calculate the energy penalty analytically. 
For simplicity, quadrature phase shift keying (QPSK) is used, and power 
allocation is not considered. 

\section{Precoding}
\subsection{Zero-Forcing Transmit Beamforming} 
We start with the conventional ZF-BF~\cite{Wiesel08}, assuming $K\leq N$. 
Let $\boldsymbol{x}_{t}=(x_{1,t},\ldots,x_{K,t})^{\mathrm{T}}$ denote the QPSK 
data symbol vector with unit power in time slot~$t$. The transmit vector 
$\boldsymbol{u}_{t}$ is linear-precoded as follows:  
\begin{equation} \label{ZF-BF} 
\boldsymbol{u}_{t} 
= \boldsymbol{H}^{\mathrm{H}}(\boldsymbol{H}\boldsymbol{H}^{\mathrm{H}})^{-1}
\boldsymbol{x}_{t}. 
\end{equation} 
Substituting (\ref{ZF-BF}) into (\ref{MIMO-BC}) implies that 
inter-user interference is eliminated completely. More precisely, the 
MIMO-BC~(\ref{MIMO-BC}) is decomposed into single-user Gaussian channels with 
receive SNR~$1/(\mathcal{E}N_{0})$. The drawback of ZF-BF is that the energy 
penalty~(\ref{energy_penalty}) in $T_{\mathrm{c}}\rightarrow\infty$ 
\begin{equation} \label{energy_penalty_ZF}
\frac{\mathcal{E}}{K} =
\frac{1}{KT_{\mathrm{c}}}\sum_{t=0}^{T_{\mathrm{c}}-1}
\boldsymbol{x}_{t}^{\mathrm{H}}(\boldsymbol{H}\boldsymbol{H}^{\mathrm{H}})^{-1}
\boldsymbol{x}_{t}  
\rightarrow
\frac{1}{K}\mathrm{Tr}\left\{
 (\boldsymbol{H}\boldsymbol{H}^{\mathrm{H}})^{-1}
\right\} 
\end{equation}
diverges as $\alpha=K/N\rightarrow1$~\cite{Tulino04}.  
Consequently, the receive SNR tends to zero as $\alpha\rightarrow1$. 
As a solution to circumventing the divergence of the energy penalty, VP has 
been considered. 

\subsection{Vector Precoding} 
In VP with ZF-BF, each data symbol $x_{k,t}$ is modified to take values in a 
relaxed alphabet $\mathcal{X}_{x_{k,t}}\subset\mathbb{C}$. The relaxed 
alphabets for different data symbols must be disjoint, i.e.\ 
$\mathcal{X}_{x}\cap\mathcal{X}_{\tilde{x}}=\emptyset$ for all $x\neq\tilde{x}$. 
Since information is conveyed by the relaxed alphabet 
$\mathcal{X}_{x_{k,t}}$, the receiver detects the relaxed alphabet 
$\mathcal{X}_{x_{k,t}}$. See \cite{Hochwald05} for the details.    
If one uses the vector $\tilde{\boldsymbol{x}}_{t}$ to minimize the energy 
penalty~(\ref{energy_penalty}) as the modified vector, the transmit vector 
is given by 
\begin{equation}
\boldsymbol{u}_{t} 
= \boldsymbol{H}^{\mathrm{H}}(\boldsymbol{H}\boldsymbol{H}^{\mathrm{H}})^{-1}
\tilde{\boldsymbol{x}}_{t}, 
\end{equation}
with 
\begin{equation} \label{VP} 
\tilde{\boldsymbol{x}}_{t} 
= \argmin_{\tilde{\boldsymbol{x}}_{t}\in\prod_{k=1}^{K}\mathcal{X}_{x_{k,t}}}
\tilde{\boldsymbol{x}}_{t}^{\mathrm{H}}
(\boldsymbol{H}\boldsymbol{H}^{\mathrm{H}})^{-1}
\tilde{\boldsymbol{x}}_{t}. 
\end{equation}
Note that the vector~(\ref{VP}) to minimize each instantaneous 
power $\|\boldsymbol{u}_{t}\|^{2}$ minimizes the energy 
penalty~(\ref{energy_penalty}) for any $T_{\mathrm{c}}$. 

\begin{example}[LVP]
In LVP~\cite{Hochwald05}, two-dimensional (one-complex-dimensional) square 
lattices are used as the relaxed alphabets, 
\begin{equation} \label{LVP} 
\mathcal{X}_{x} = \frac{4}{\sqrt{2}}\mathbb{Z}+\Re[x] 
+ \mathrm{i}\left(
 \frac{4}{\sqrt{2}}\mathbb{Z}+\Im[x]
\right), 
\end{equation}
for $\Re[x], \Im[x]\in\{\pm1/\sqrt{2}\}$. It is infeasible in terms of the 
complexity to find the optimal vector~(\ref{VP}) for LVP. 
\end{example} 
\begin{example}[CVP]
In CVP~\cite{Mueller08}, the original alphabets are relaxed to continuous 
disjoint alphabets, 
\begin{equation} \label{CVP} 
\mathcal{X}_{x} 
= \tilde{\mathcal{X}}_{\Re[x]} + \mathrm{i}\tilde{\mathcal{X}}_{\Im[x]}, 
\end{equation}
with 
\begin{equation}
\tilde{\mathcal{X}}_{x} 
= \left\{
\begin{array}{ll}
[x,\infty) & \hbox{for $x=1/\sqrt{2}$} \\  
(-\infty,x] & \hbox{for $x=-1/\sqrt{2}$.}  
\end{array}
\right. 
\end{equation}
The minimization~(\ref{VP}) for CVP reduces to a convex optimization 
problem, so that an efficient algorithm can be used to solve 
(\ref{VP}). However, CVP might be still hard to implement 
since the convex optimization needs to be solved every time slot.  
\end{example}

The point of VP is that the modified vector $\tilde{\boldsymbol{x}}_{t}$ 
depends on the channel matrix $\boldsymbol{H}$. Consequently, the energy 
penalty $T_{\mathrm{c}}^{-1}\sum_{t=0}^{T_{\mathrm{c}}-1}
\tilde{\boldsymbol{x}}_{t}^{\mathrm{H}}
(\boldsymbol{H}\boldsymbol{H}^{\mathrm{H}})^{-1}\tilde{\boldsymbol{x}}_{t}$ 
for VP never tends to the right-hand side (RHS) of (\ref{energy_penalty_ZF}) 
in $T_{\mathrm{c}}\rightarrow\infty$. In fact, the energy penalty for VP was 
shown to be bounded in the limit $\alpha\rightarrow1$ after taking the 
large-system limit~\cite{Zaidel12}.

\begin{figure*}[!t]
% ensure that we have normalsize text
\normalsize
% Store the current equation number.
\setcounter{MYtempeqncnt}{\value{equation}}
% Set the equation number to one less than the one
% desired for the first equation here.
% The value here will have to changed if equations
% are added or removed prior to the place these
% equations are referenced in the main text.
\setcounter{equation}{10}
\begin{equation} \label{US_VP} 
(\mathcal{K}_{i},\{\tilde{\boldsymbol{x}}_{\mathcal{K}_{i},t}\}) 
= \argmin_{\mathcal{K}_{i}\subset\mathcal{K}:|\mathcal{K}_{i}|=\tilde{K}}
\argmin_{\{\tilde{\boldsymbol{x}}_{\mathcal{K}_{i},t}
\in\prod_{k\in\mathcal{K}_{i}}\mathcal{X}_{x_{k,t}}:t\}}
\frac{1}{T}\sum_{t=iT}^{(i+1)T-1}
\tilde{\boldsymbol{x}}_{\mathcal{K}_{i},t}^{\mathrm{H}}
(\boldsymbol{H}_{\mathcal{K}_{i}}\boldsymbol{H}_{\mathcal{K}_{i}}^{\mathrm{H}})^{-1}
\tilde{\boldsymbol{x}}_{\mathcal{K}_{i},t}.  
\end{equation}
% Restore the current equation number.
\setcounter{equation}{\value{MYtempeqncnt}}
% IEEE uses as a separator
\hrulefill
% The spacer can be tweaked to stop underfull vboxes.
\vspace*{4pt}
\end{figure*}

\subsection{Joint US and VP} \label{sec34} 
US-VP is performed every $T$ time slots. The block 
length $T$ should not be confused with the coherence time $T_{\mathrm{c}}$. 
We write the set of selected users in the $i$th block of US-VP as 
$\mathcal{K}_{i}$ with $|\mathcal{K}_{i}|=\tilde{K}$, for $i=0,1,\ldots$. 
The base station sends the QPSK data symbols $\{x_{k,iT+t}:t=0,\ldots,T-1\}$ 
to the selected user~$k\in\mathcal{K}_{i}$ in block~$i$. 
The transmit vector $\boldsymbol{u}_{t}$ ($t\in[iT,(i+1)T-1]$) in block~$i$ 
is generated as 
\begin{equation} \label{ZF} 
\boldsymbol{u}_{t} = \boldsymbol{H}_{\mathcal{K}_{i}}^{\mathrm{H}}
(\boldsymbol{H}_{\mathcal{K}_{i}}
\boldsymbol{H}_{\mathcal{K}_{i}}^{\mathrm{H}})^{-1}
\tilde{\boldsymbol{x}}_{\mathcal{K}_{i},t},  
\end{equation}
where the set of selected users $\mathcal{K}_{i}\subset
\mathcal{K}=\{1,\ldots,K\}$ and 
the modified vectors $\{\tilde{\boldsymbol{x}}_{\mathcal{K}_{i},t}\in
\prod_{k\in\mathcal{K}_{i}}\mathcal{X}_{x_{k,t}}\}$ minimize the energy 
penalty~(\ref{energy_penalty}): They are given by (\ref{US_VP}) at the top of 
this page.   
\setcounter{equation}{11} 

It is difficult to solve the minimization~(\ref{US_VP}) 
for US-LVP and US-CVP, which are defined as US-VP with (\ref{LVP}) and 
(\ref{CVP}), respectively. Instead, we focus on DD-US. 

\begin{example}[Data-Dependent US]
DD-US is defined as US-VP with the original alphabet as the relaxed alphabet, 
i.e.\ $\mathcal{X}_{x}=\{x\}$. Since DD-US is performed every $T$ time slots, 
DD-US may be more suitable for implementation. A greedy algorithm 
for DD-US with ZF-BF proposed in \cite{Takeuchi121} allows us to solve the 
minimization~(\ref{US_VP}) efficiently and approximately. 
\end{example}

Let us discuss the relationship between DD-US and conventional US, the latter 
of which selects a subset of users $\mathcal{K}_{\mathrm{c}}\subset\mathcal{K}$ 
to minimize the energy penalty $\mathrm{Tr}\{
(\boldsymbol{H}_{\mathcal{K}_{\mathrm{c}}}
\boldsymbol{H}_{\mathcal{K}_{\mathrm{c}}}^{\mathrm{H}})^{-1}\}$. The 
energy penalty of the conventional US is obviously larger than 
that of DD-US for any $T$. In DD-US, the set of selected users 
$\mathcal{K}_{i}$ is determined on the basis of an appropriate 
tradeoff~(\ref{US_VP}) between the orthogonality of the channel row vectors 
and the direction of the data symbols. The performance of DD-US degrades 
as $T$ increases, since it becomes difficult to select those data symbols with 
{\em good} direction. Thus, the energy penalty of DD-US in 
$T\rightarrow\infty$ can be regarded as a lower bound on the 
energy penalty for the conventional US. 

Each user has to detect whether he/she has been selected in each block of US.  
In order for each user to blind-detect it, the data symbols for non-selected 
users should be discarded at the {\em transmitter} side~\cite{Takeuchi121}. 
Substituting (\ref{ZF}) into (\ref{MIMO-BC}) yields 
\begin{equation} \label{equivalent_channel} 
y_{k,t} = \frac{1}{\sqrt{\mathcal{E}}}\left\{
 s_{k,i}\tilde{x}_{k,t} + (1-s_{k,i})\vec{\boldsymbol{h}}_{k}\boldsymbol{u}_{t}  
\right\} + n_{k,t}, 
\end{equation}
for any $k$. In (\ref{equivalent_channel}), 
$\tilde{x}_{k,t}\in\mathcal{X}_{x_{k,t}}$ denotes the modified data symbol 
corresponding to the original data symbol $x_{k,t}$. The variable 
$s_{k,i}\in\{0,1\}$ indicating whether user~$k$ has been selected in block~$i$ 
is defined as 
\begin{equation} \label{indicator}
s_{k,i} = \left\{
\begin{array}{ll}
1 & k\in\mathcal{K}_{i} \\ 
0 & k\notin\mathcal{K}_{i}.  
\end{array}
\right.
\end{equation}
Furthermore, $\vec{\boldsymbol{h}}_{k}\in\mathbb{C}^{1\times N}$ denotes the 
$k$th row vector of the channel matrix $\boldsymbol{H}$. 
Note that the indices $t$ of $y_{k,t}$ and $\tilde{x}_{k,t}$ in 
(\ref{equivalent_channel}) are identical to 
each other, since the data symbols for the non-selected users 
$k\notin\mathcal{K}_{i}$ have been discarded. This simplifies the detection 
of (\ref{indicator})~\cite{Takeuchi121}.   

It is easy for user~$k$ to blind-detect {\em one} variable $s_{k,i}$ from 
the $T$ observations $\{y_{k,t}\}$ in each block. Using the decision-feedback 
of $\tilde{x}_{k,t}$ from the decoder improve the accuracy of 
detection~\cite{Takeuchi121}. In order to reduce the energy penalty, 
small $T$ should be used. On the other hand, too small $T$ makes it 
difficult to detect. As one option, dozens of time slots should be used as 
the block length $T$. For example, the energy loss due to detection errors is 
at most 0.2--0.5~dB for $T=16$~\cite{Takeuchi121}.

\section{Main Results}
The replica method is used to analyze the energy penalty for US-VP 
in the large-system limit, where $N$, $K$, and $\tilde{K}$ tend to infinity 
with the ratios $\alpha=K/N$ and $\kappa=\tilde{K}/K$ kept constant. 
The energy penalty is expected to be self-averaging in the 
large-system limit: It converges in probability (or almost surely) to 
the expected one in the large-system limit. Thus, we focus on the average 
energy penalty.  

We consider the assumptions of replica symmetry (RS) and of 1-step 
replica symmetry breaking (1RSB)~\cite{Mezard87,Nishimori01}. Roughly 
speaking, the RS assumption corresponds to the assumption that the solution 
to the minimization~(\ref{US_VP}) is unique. 
On the other hand, 1RSB is the simplest 
assumption for the case in which there are many solutions. It is empirically 
known that the 1RSB assumption can provide a good approximation for the 
energy penalty~\cite{Zaidel12}. 

Without loss of generality, we focus on the first block of US, i.e.\ 
$t=0,\ldots,T-1$. Before presenting the main results, we summarize several 
definitions. Let us define a random variable $E_{k}(q)$ as 
\begin{equation} \label{E} 
E_{k}(q) = \frac{1}{T}\sum_{t=0}^{T-1}
\min_{\tilde{x}_{k,t}\in\mathcal{X}_{x_{k,t}}}|\tilde{x}_{k,t} - \sqrt{q}z_{k,t}|^{2},  
\end{equation} 
with $z_{k,t}\sim\mathcal{CN}(0,1)$. We write the 
cumulative distribution function $\mathrm{Prob}(E_{k}(q)\leq x)$ for the 
random variable~(\ref{E}) and its inverse function as $F_{T}(x;q)$ and 
$F_{T}^{-1}(x;q)$, respectively. We define two quantities 
$\mu_{\kappa,T}(q)$ and $\sigma_{\kappa,T}^{2}(q)$ as 
\begin{equation} \label{mu}
\mu_{\kappa,T}(q) = \int_{0}^{\kappa}F_{T}^{-1}(x;q)dx, 
\end{equation}
\begin{IEEEeqnarray}{r}
\sigma_{\kappa,T}^{2}(q) = 
\int_{0}^{F_{T}^{-1}(\kappa;q)}\int_{0}^{F_{T}^{-1}(\kappa;q)} 
[F_{T}(\min(x,y);q) \nonumber \\ 
 - F_{T}(x;q)F_{T}(y;q) ]dxdy, \label{sigma2} 
\end{IEEEeqnarray}
respectively. These quantities are associated with the mean and variance of 
\begin{equation}
E(q) = \frac{1}{K}\sum_{k=1}^{\tilde{K}}E_{(k)}(q), 
\end{equation}
where $\{E_{(k)}(q)\}$ are the order statistics of (\ref{E}), i.e.\ 
$E_{(1)}(q)\leq\cdots\leq E_{(K)}(q)$~\cite{Stigler74}.  

\begin{proposition} \label{proposition1} 
Under the RS assumption, the average energy penalty per selected user 
$\mathbb{E}[\mathcal{E}]/\tilde{K}$ for US-VP converges to 
$q_{0}/(\alpha\kappa)$ in the large-system limit, 
which is the solution to the fixed-point equation
\begin{equation} \label{RS} 
q_{0} = \alpha\mu_{\kappa,T}(q_{0}).
\end{equation}
\end{proposition}
\begin{proposition} \label{proposition2} 
Under the 1RSB assumption, the average energy penalty per selected user 
$\mathbb{E}[\mathcal{E}]/\tilde{K}$ for US-VP converges to 
$q_{1}/(\alpha\kappa)$ in the large-system limit, which satisfies the 
coupled fixed-point equations 
\begin{equation} \label{1RSB_1}
g(q_{1},\chi)=0, 
\end{equation}
\begin{equation} \label{1RSB_2} 
\frac{\partial}{\partial\chi}g(q_{1},\chi)=0. 
\end{equation}
for some $\chi>0$, with 
\begin{equation}
g(q_{1},\chi) = \ln\left(
 1 + \frac{q_{1}}{\chi}
\right)
- \frac{\alpha}{\chi}\left(
 \mu_{\kappa,T}(q_{1}) 
 - \frac{T\sigma_{\kappa,T}^{2}(q_{1})}{2\chi}
\right). 
\end{equation}
\end{proposition}

See \cite[Appendices~C and D]{Takeuchi122} for the details of the derivations. 
The central limit theorem implies that the random variable~(\ref{E}) 
converges in law to a Gaussian random in $T\rightarrow\infty$. 
It is straightforward to find that (\ref{mu}) reduces to 
\begin{equation}
\lim_{T\rightarrow\infty}\mu_{\kappa,T}(q) 
= \kappa\mathbb{E}\left[
 \min_{\tilde{x}_{t}\in\mathcal{X}_{x_{1,t}}}|\tilde{x}_{t} - \sqrt{q}z_{t}|^{2} 
\right]. 
\end{equation}
It is worth noting that the energy penalty of DD-US under the RS assumption 
is explicitly given by  
\begin{equation} \label{energy_penalty_inf} 
\frac{\mathbb{E}[\mathcal{E}]}{\tilde{K}} \rightarrow \frac{1}{1-\alpha\kappa} 
= \frac{1}{1 - \tilde{K}/N}, 
\end{equation}
as the block length $T$ tends to infinity. 
The energy penalty~(\ref{energy_penalty_inf}) 
under the RS assumption is equal to that for ZF-BF with random US (RUS), 
in which $\tilde{K}$ users are selected uniformly and randomly. Similarly, 
the energy penalty for US-CVP under the RS assumption is also equal to 
that for RUS and CVP (RUS-CVP) in $T\rightarrow\infty$. Since the 
energy penalty for DD-US in $T\rightarrow\infty$ is a lower bound on that for 
conventional US with ZF-BF, one may conclude that conventional US 
makes no sense in the large-system limit. However, we cannot reach this  
conclusion only from those observations. The RS solutions are 
approximations for the true energy penalty in the large-system limit. 
In order to investigate whether the conclusion is correct, the assumption of 
higher-step RSB should be considered. 

\section{Numerical Results} 
DD-US is compared to US-CVP, ZF-BF with RUS, and RUS-CVP~\cite{Mueller08} 
in terms of the average energy penalty. Note that the block length $T$ 
is kept finite, while the coherence time $T_{\mathrm{c}}$ is implicitly 
assumed to tend to infinity. We found that 
Propositions~\ref{proposition1} and~\ref{proposition2} for US-LVP provide 
unreliable approximations for the energy penalty, so that US-LVP is not 
plotted. The assumption of higher-step RSB is required to obtain a good 
approximation for US-LVP. 

\begin{figure}[t]
\begin{center}
\includegraphics[width=\hsize]{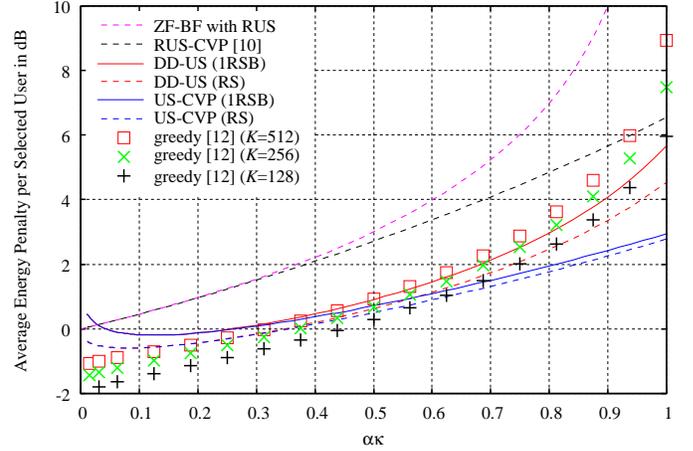}
\end{center}
\caption{
$\mathbb{E}[\mathcal{E}]/\tilde{K}$ versus $\alpha\kappa=\tilde{K}/N$ 
for $\alpha=4$ and $T=20$. 
}
\label{fig1} 
\end{figure}

Figure~\ref{fig1} shows the average energy penalties per selected user of 
the four schemes for $T=20$ in the large-system limit. The RS and 1RSB 
solutions are plotted by dashed and solid lines, respectively. The energy 
penalties for a greedy algorithm of DD-US~\cite{Takeuchi121} are also shown 
for $K=128, 256, 512$. The 1RSB assumptions are obviously unreliable for 
small $\alpha\kappa$, since the energy penalties are larger than those  
for ZF-BF with RUS and RUS-CVP, which correspond to upper bounds.   
We can observe four results: First, the gap between 
the RS and 1RSB solutions for US-CVP is small for moderate-to-large 
$\alpha\kappa$, while the gap for DS-US is for moderate $\alpha\kappa$. 
Secondly, as the system size increases, the energy penalties for the greedy 
algorithm of DD-US~\cite{Takeuchi121} get closer from {\em below} to the RS 
solution for small $\alpha\kappa$ and to the 1RSB solution for moderate 
$\alpha\kappa$, respectively. These results imply that the RS and 1RSB 
solutions for DD-US can provide acceptable approximations for small  
$\alpha\kappa$ and for moderate $\alpha\kappa$, respectively. Thirdly, DD-US 
achieves almost the same energy penalty as US-CVP for low system loads. This 
result can be understood as follows: $q$ in (\ref{E}) is small for low system 
loads, so that the magnitude of $\sqrt{q}\Re[z_{t}]$ ($\sqrt{q}\Im[z_{t}]$) is 
smaller than the magnitude of the data symbol with high probability. Thus, 
the continuous relaxation of the alphabet~(\ref{CVP}) makes no sense in this 
region of the system load. Finally, we find that DD-US outperforms RUS-CVP 
in terms of the energy penalty except for high $\alpha\kappa$. 
For $\alpha\kappa=0.5$, DD-US can provide a performance gain of $1.8$~dB, 
compared to RUS-CVP, which seems to be larger than the energy loss due to 
the detection error at the receiver~\cite{Takeuchi121}, noted in the end of 
Section~\ref{sec34}.  
Note that the energy penalty for RUS-CVP gets closer from {\em above} to 
the asymptotic one as the system size increases~\cite{Zaidel12}. Thus, 
the performance gap between DD-US and RUS-CVP should be larger for 
finite-sized systems. 

\begin{figure}[t]
\begin{center}
\includegraphics[width=\hsize]{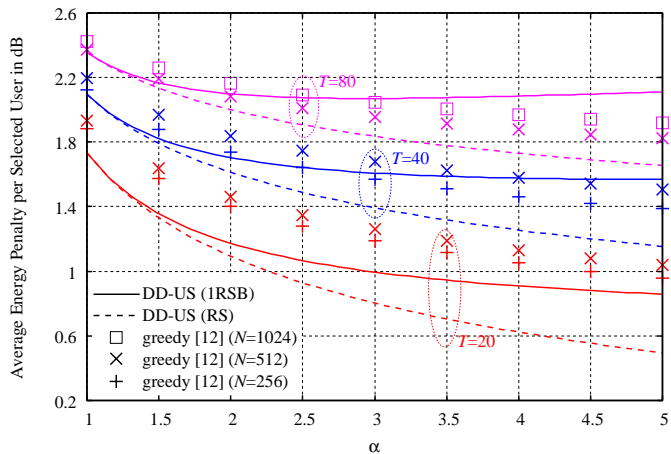}
\end{center}
\caption{
$\mathbb{E}[\mathcal{E}]/\tilde{K}$ versus $\alpha$ for $\alpha\kappa=0.5$. 
}
\label{fig2} 
\end{figure}

We next assess the accuracy of the approximations based on the RS and 
1RSB assumptions for DD-US. Figure~\ref{fig2} shows the average energy 
penalty per selected user versus $\alpha$ for fixed $\alpha\kappa=0.5$. For 
comparison, the energy penalties for the greedy algorithm of 
DD-US~\cite{Takeuchi121} are also plotted. 
For small $\alpha$, the RS and 1RSB solutions are 
indistinguishable from each other, so that they should provide an accurate 
approximation of the true energy penalty for small $\alpha$. The gaps between 
the analytical results and the numerical simulations for small $\alpha$ should 
be due to the suboptimality of the greedy algorithm. 
The 1RSB solution for $T=80$ exhibits strange behavior: 
The energy penalty must be a monotonically decreasing function of $\alpha$, 
since large $\alpha$ implies large multiuser diversity. However, 
the energy penalty {\em increases} with the {\em increase} of $\alpha$ 
for large $\alpha$. 

\section{Conclusions}
The energy penalties of DD-US and US-CVP for the MIMO-BC have been evaluated 
in the large-system limit under the RS and 1RSB assumptions. We found 
four observations: First, the RS and 1RSB assumptions for DD-US can provide 
acceptable approximations for low and moderate 
system loads, respectively. Secondly, DD-US outperforms RUS-CVP in terms of 
the energy penalty for low-to-moderate system loads. Thirdly, the asymptotic 
energy penalty of DD-US is indistinguishable from that of US-CVP for low 
system loads. Finally, a greedy algorithm of DD-US proposed in 
\cite{Takeuchi121} can achieve nearly optimal performance for low-to-moderate 
system loads. These results imply that DD-US can provide a good tradeoff 
between the performance and the complexity for low-to-moderate system loads. 

%In this paper, we have focused on the energy penalty. The achievable sum 
%rate should be investigated to make more detailed comparisons. Furthermore, 
%Gaussian signaling should be considered to increase the achievable sum rate. 
As another method for reducing the energy penalty, 
it is important to investigate regularized ZF-BF or minimum mean-squared error 
(MMSE) precoding. We leave this analysis as future work. 

\section*{Acknowledgment}
The work of K.~Takeuchi was in part supported by the JSPS Institutional 
Program for Young Researcher Overseas Visits and by the Grant-in-Aid for 
Young Scientists~(B) (No.~23760329) from MEXT, Japan.    

\balance

\bibliographystyle{IEEEtran}
\bibliography{IEEEabrv,kt-isita2012}

\end{document}